\documentclass[prb,aps,twocolumn,superscriptaddress]{revtex4-2}

\usepackage{graphicx,amsfonts,times,bm,amsmath,verbatim,xcolor,array,subcaption}
\graphicspath{{figs/}}

\usepackage{float}
\usepackage{hyperref}
\hypersetup{
	colorlinks=true,final=true,
	linkcolor=blue,
	citecolor=blue,
	filecolor=blue,
	urlcolor=blue,
}

\begin{document}
	
	\title{ Giant anomalous thermal Hall effect in tilted type-I magnetic Weyl semimetal Co$_3$Sn$_2$S$_2$}
	
	\author{Abhirup Roy Karmakar}
	\email{abhirup.phy@iitkgp.ac.in}
	\affiliation{Department of Physics, Indian Institute of Technology Kharagpur, W.B. 721302, India}
	
	\author{S. Nandy}
	\affiliation{Department of Physics, University of Virginia, Charlottesville, VA 22904, USA}
	
	\author{A. Taraphder}
	\email{arghya@phy.iitkgp.ac.in}
	\affiliation{Department of Physics, Indian Institute of Technology Kharagpur, W.B. 721302, India}
	
	\author{G. P. Das}
	\email{gour.das@tcgcrest.org}
	\affiliation{Research Institute for Sustainable Energy (RISE), TCG Centres for Research and Education in Science and Technology, Sector V, Salt Lake, Kolkata 700091, India}
	
	\date{\today}
	
\begin{abstract}
	The recent discovery of magnetic Weyl semimetal Co$ _3 $Sn$ _2 $S$ _2 $ opens up new avenues for research into the interactions between topological orders, magnetism, and electronic correlations. Motivated by the observations of large anomalous Hall effect because of large Berry curvature, we investigate another Berry curvature-induced phenomenon, the anomalous thermal Hall effect in Co$ _3 $Sn$ _2 $S$ _2$. We study it with and without strain, using a Wannier tight-binding Hamiltonian derived from first principles density functional theory calculations. We first identify this material as a tilted type-I Weyl semimetal based on the band structure calculation. Within the quasi-classical framework of Boltzmann transport theory, a giant anomalous thermal Hall signal appears due to the presence of large Berry curvature. Surprisingly, the thermal Hall current changes and even undergoes a sign-reversal upon varying the chemical potential. Furthermore, applying about 13 GPa stress, an enhancement as large as 33\% in the conductivity is observed; however, the tilt vanishes along the path connecting the Weyl nodes. In addition, we have confirmed the validity of the Wiedemann-Franz law in this system for anomalous transports. We propose specific observable signatures that can be directly tested in experiments. \\ \\
	DOI: \href{https://doi.org/10.1103/PhysRevB.106.245133}{10.1103/PhysRevB.106.245133}
\end{abstract}
	
	\maketitle
	
\section{Introduction}\label{sec:introduction}
	
	Topologically protected gapped and gapless materials have attracted immense attention lately due to their unique properties \cite{chiu_classification_2016,armitage_weyl_2018}. In line with band structure terminology, they are known as topological insulators \cite{hasan_colloquium_2010,qi_topological_2011} and topological semimetals respectively \cite{armitage_weyl_2018,yan_topological_2017}. The three-dimensional (3D) realizations of topological semimetals, which include Dirac and Weyl semimetals (DSM, WSM), have been classified in accordance with the symmetries they possess as well as the nature of band dispersion near the Fermi level \cite{armitage_weyl_2018,yan_topological_2017}.
	
	In the case of a WSM, the Weyl nodes, which are earmarked by their locations in momentum and energy space where the non-degenerate bands linearly touch each other, always appear in pairs with well-defined but opposite chiralities~\cite{murakami_phase_2007,murakami_tuning_2007,yang_quantum_2011, burkov_weyl_2011, wan_topological_2011, xu_chern_2011}. Each node of the pairs individually acts as the source (+ve chirality) or sink (-ve chirality) of the Berry curvature, which can be viewed as an effective magnetic field in momentum space~\cite{xiao_berry_2010}. In order to have a topological charge associated with the Weyl node, WSM has to break either the time-reversal symmetry (TRS) or the space inversion symmetry (IS) which leads to the classification of Weyl semimetals:  (i) IS broken but TR-symmetric WSM and (ii) TRS broken but inversion-symmetric WSM or magnetic WSM~\cite{burkov_weyl_2011,wan_topological_2011, xu_chern_2011, volovik_emergent_2014}. Compared to the IS broken WSM, the magnetic WSM gives a platform to study the interplay between topological orders, magnetism and electronic correlations, leading to exotic quantum states. In addition, it can also generate a true nodal WSM phase when the Fermi energy coincides with the Weyl nodes since the pair of Weyl nodes remains at the same energy to preserve IS. 
	
	Ever since their discovery, Weyl semimetals have been surprising us with their unique and fascinating transport properties, such as nontrivial Berry curvature induced anomalous Hall effect (AHE) and anomalous Nernst effect (ANE)~\cite{gorbar_anomalous_2017,burkov_anomalous_2014, 
	steiner_anomalous_2017,shekhar_anomalous_2018,
	watzman_dirac_2018,caglieris_anomalous_2018,ferreiros_anomalous_2017,roykarmakar_probing_2021}, chiral anomaly induced negative longitudinal magnetoresistance~\cite{huang_observation_2015,zyuzin_topological_2012,zhang_signatures_2016,
	moll_magnetic_2016,zeng_chiral_2022,zeng_nonlinear_2021}, planar Hall effect~\cite{ghosh_chiralitydependent_2020,nag_magnetotransport_2020,burkov_giant_2017,
	nandy_chiral_2017,singha_planar_2018,
	kumar_planar_2018,chen_planar_2018,
	sharma_revisiting_2022,yang_current_2019}, thermoelectric phenomena~\cite{lundgren_thermoelectric_2014,nandy_planar_2019,
	sharma_nernst_2016}, gyrotropic birefringence~\cite{nandy_nonreciprocal_2021}, Magnus Hall effect~\cite{das_topological_2021}, and chiral magnetic effect~\cite{nandy_chiral_2020}.
	
	These topological transport properties in WSM have been studied extensively both theoretically and experimentally. However, in the context of real materials, most of these studies are based on the IS-broken but TR-symmetric WSMs. After the first experimental discovery of WSM in TaAs in 2015~\cite{xu_discovery_2015, lv_experimental_2015}, the other WSM candidates discovered so far are non-magnetic. On the other hand, despite many proposed candidates, a direct experimental verification of magnetic WSM is challenging. In 2019, a Shandite material Co$ _3 $Sn$ _2 $S$ _2 $, in the ferromagnetic phase, has been identified as a Weyl semimetal with three pairs of Weyl points in the Brillouin zone~\cite{belopolski_discovery_2019, liu_magnetic_2019, morali_fermiarc_2019}. Due to its intrinsic magnetism, as well as a large Berry curvature, Co$ _3 $Sn$ _2 $S$ _2 $ shows large anomalous Hall effect, anomalous Nernst effect, planar Hall effect and chiral anomaly-induced magnetotransport phenomena, all of which have been studied extensively \cite{yang_giant_2020,yang_fieldmodulated_2020,liu_giant_2018,li_epitaxial_2020,okamura_giant_2020,thakur_intrinsic_2020,wang_large_2018,shama_observation_2020,chen_pressuretunable_2019,tanaka_topological_2020,guin_zero_2019}. However, another important effect that is gaining attention lately, viz. the anomalous thermal Hall effect (ATHE), has not been explored in this particular class of material as yet. ATHE, in contrast to conventional heat conduction, causes the generation of heat current in the direction transverse to the applied temperature gradient in the absence of any external magnetic field. It is also known as the anomalous Righi-Leduc effect and can serve as an additional source of dissipationless heat current  \cite{zhang_determining_2000, onose_lorenz_2008, li_anomalous_2017}. Similar to other anomalous transports, this one is caused by the non-trivial topology of the bands, with Berry curvature playing the role of the magnetic field. The finite results of the thermal Hall effect represent both the system's topology and the anomalous scattering of quasiparticle excitations. ATHE has been utilized as a potent instrument for investigating charge-neutral quasiparticles in insulating quantum materials \cite{hirschberger_thermal_2015, hirschberger_large_2015, doki_spin_2018,ideue_giant_2017}. This particular transport phenomenon may also be effective in the directional control of heat currents.
	
	In this paper, we investigate the anomalous thermal Hall effect of the magnetic WSM Co$ _3 $Sn$ _2 $S$ _2 $ in detail. We start by studying the electronic structure of this material and obtain a Wannier tight-binding model from the first principles calculation. We identify this material as a type-I Weyl semimetal with tilted cones for the first time using DFT calculations which agrees with very recent experimental findings \cite{jiang_chiralitydependent_2021,jiang_antisymmetric_2022}. Then we calculate the Berry curvature which exhibits very high values at some specific points in the Brillouin zone, namely the Weyl nodes. Interestingly, the anomalous thermal Hall conductivity (ATHC) obtained by integrating the Berry curvature comes out to be very large without the presence of any external field when compared with previous observations \cite{hwang_topological_2020,li_topological_2021,ideue_giant_2017,zhang_anomalous_2021}. 
	The result is validated by confirming that it satisfies the Wiedemann-Franz law quite accurately at low temperature. However, at higher temperatures the conductivity loses its linear relationship with temperature. By plotting ATHC over a range of chemical potentials we find that the heat current can be tuned as well as reversed by changing the chemical potential, although the maximum value is attained at its original Fermi level. Furthermore, we study the effect of external pressure on ATHC by applying uniaxial compressive stress on the crystal along different directions. We find that the application of stress along the direction perpendicular to the kagome-plane drags the Weyl points closer to the Fermi level which in turn enhances the transverse heat current. We also observe that the tilt along the connecting line between the Weyl nodes decreases to zero upon applying stress. However, stresses along lateral directions did not improve the conductivity and, therefore, we only focus our attention to the \textit{z}-axis compression.
	
	The remainder of this paper is organized as follows: in Sec. \ref{sec:matherial_and_method} we discuss the crystal structure, underlying symmetries and magnetic moments of Co$ _3 $Sn$ _2 $S$ _2$. Then we elaborate the computational methods  employed throughout the work. After that we study its electronic band structure, density of states, as well as Fermi surface in Sec. \ref{sec:electronic_structure}. In Sec. \ref{sec:athe} we investigate ATHE and study its behavior with chemical potential and temperature. Then in Sec. \ref{sec:strain} we discuss the effect of stress on the system and show how it can be used to enhance the ATHC. And finally, we summarize our work, the results obtained and possible future directions in Sec. \ref{sec:summary}.

\section{Material and Method}\label{sec:matherial_and_method}
	
	\begin{figure}[!hbt]
		\includegraphics[width=4.3cm]{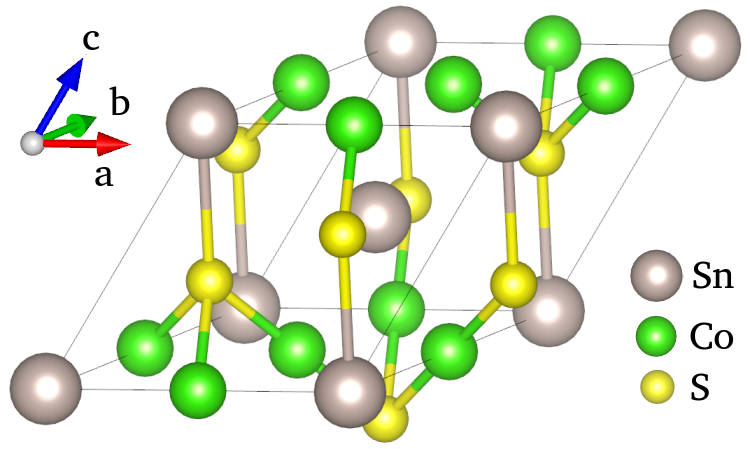}  \hfill
		\includegraphics[width=3.7cm]{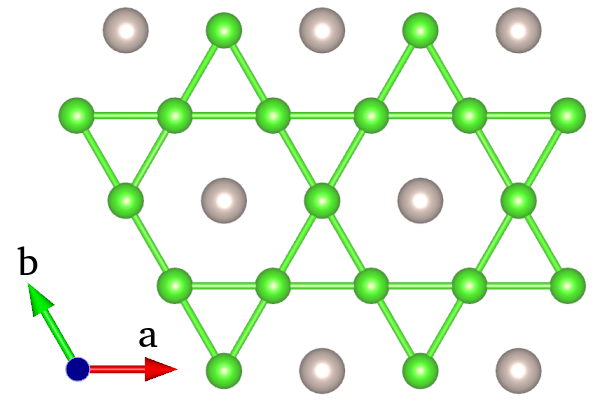}
		\caption{\textbf{Crystal structure of Co$_{\bf3}$Sn$_{\bf2}$S$_{\bf2}$.} The left part of the figure shows the rhombohedral primitive unit cell of space group \textit{R-3m}. The cobalt and tin atoms form a ferromagnetic kagome lattice which is shown at the right.} 
		\label{fig:crystal_structure}
	\end{figure}
	The shandite material Co$ _3 $Sn$ _2 $S$ _2 $ crystallizes in a rhombohedral structure with a layer of quasi-2D Co$ _3 $Sn sandwiched between S and Sn atoms \cite{weihrich_halbantiperowskite_2005,weihrich_half_2006}. The unit cell falls under the space group \textit{R-3m} (no. 166) with a lattice parameter of 5.38 \AA $ $ and consists of a mirror plane ($ M_{010} $), inversion symmetry and $C_{3z} $ rotational symmetry. The material possesses a strong ferromagnetism originating from the Cobalt atoms which sit on a kagome lattice in the \textit{ab} plane in the hexagonal representation as shown in Fig. \ref{fig:crystal_structure}. The magnetic moment is 0.29 and 0.31 $ \mu_B $/Co from neutron-diffraction \cite{vaqueiro_powder_2009} and magnetization measurement \cite{schnelle_ferromagnetic_2013} respectively. The measurements further show that the magnetic moments are directed along the \textit{c}-axis with a Curie temperature ($ T_c $) 177K. Electronic structure of also reveals that it has a half-metallic nature with a gap in the minority spin channel \cite{holder_photoemission_2009,dedkov_electronic_2008}.

	We have performed first principles density functional theory (DFT) calculations using Vienna Ab initio Simulation Package (VASP) \cite{kresse_efficient_1996} in the projector augmented wave (PAW) approximation. Generalized gradient approximation (GGA) was considered for exchange-correlation functional in the Perdew-Burke-Ernzerhof (PBE) scheme \cite{perdew_generalized_1996}. A 15$\times$15$\times$15 Monk-horst grid was taken to fill the 3D Brillouin zone. From the basic DFT calculation we obtained the value of the magnetic moment of around 0.33 $ \mu_B $/Co, which agrees well with the experiments and then studied the electronic structure as described in Sec. \ref{sec:electronic_structure}. We projected the Bloch wave functions into the maximally localized Wannier functions (MLWFs) and obtained Hamiltonian matrix elements between the MLWFs with the help of Wannier90 package \cite{pizzi_wannier90_2020}. As we shall see later, the \textit{d}-orbital of Co and \textit{p}-orbitals of Sn and S have the largest contribution near Fermi level and, therefore, they were chosen as the projections for the Wannierization. In order for a thorough investigation of the band structure and calculation of ATHC, we derived a Hamiltonian using the Wannier tight-binding model with 54 bands from the matrix elements \cite{pizzi_wannier90_2020, gresch_automated_2018}. All other numerical calculations were performed in Python programming language.

\section{Electronic structure}\label{sec:electronic_structure}

		\begin{figure*}[!hbt]
			\centering
			\begin{subfigure}{0.45\textwidth}
				\caption{}
				\label{sfig:bands}
				\includegraphics[width=\textwidth]{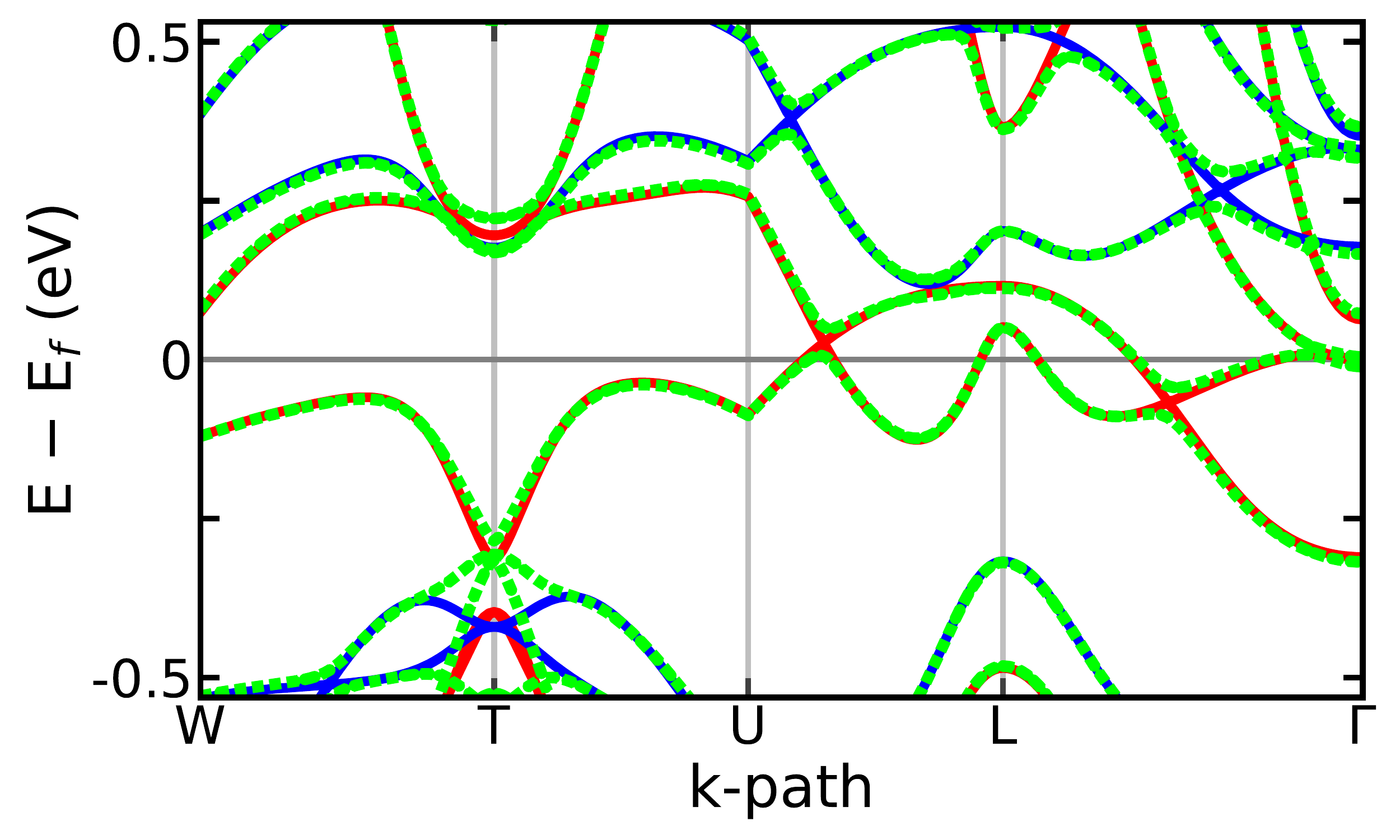}
			\end{subfigure} \hfill
			\begin{subfigure}{0.45\textwidth}
				\caption{}
				\label{sfig:pdos}
				\includegraphics[width=\textwidth]{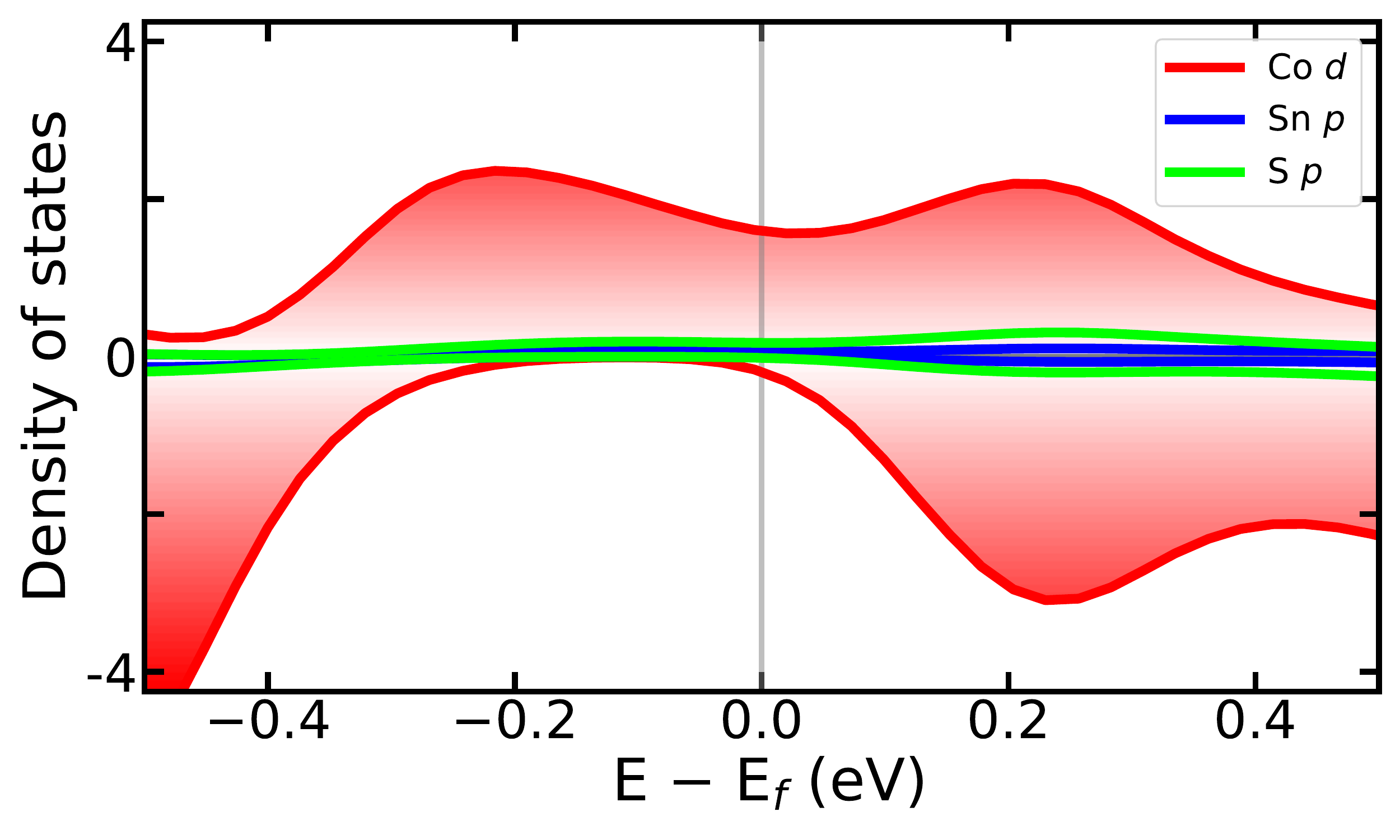}
			\end{subfigure}
			\begin{subfigure}{0.24\textwidth}
				\caption{}
				\label{sfig:brillouin_zone}
				\includegraphics[width=\textwidth]{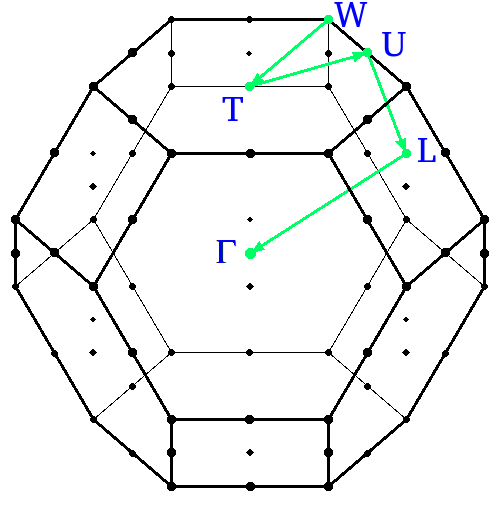}
			\end{subfigure} \hspace{1cm}
			\begin{subfigure}{0.25\textwidth}
				\caption{}
				\label{sfig:fs_vb}
				\includegraphics[width=\textwidth]{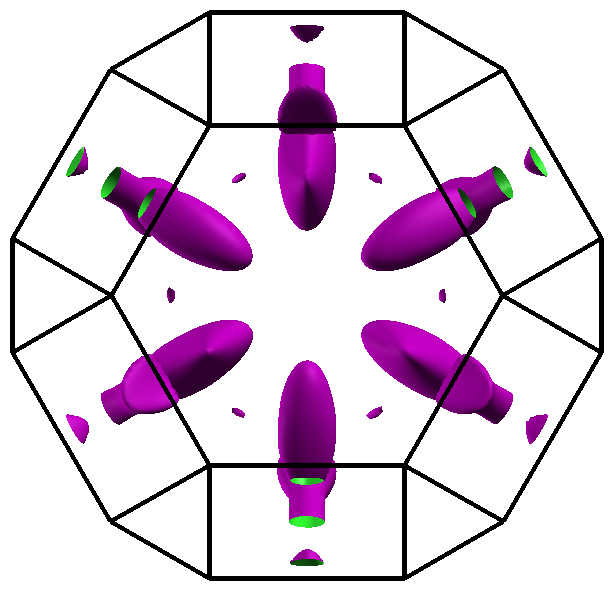}
			\end{subfigure} \hspace{1cm}
			\begin{subfigure}{0.25\textwidth}
				\caption{}
				\label{sfig:fs_cb}
				\includegraphics[width=\textwidth]{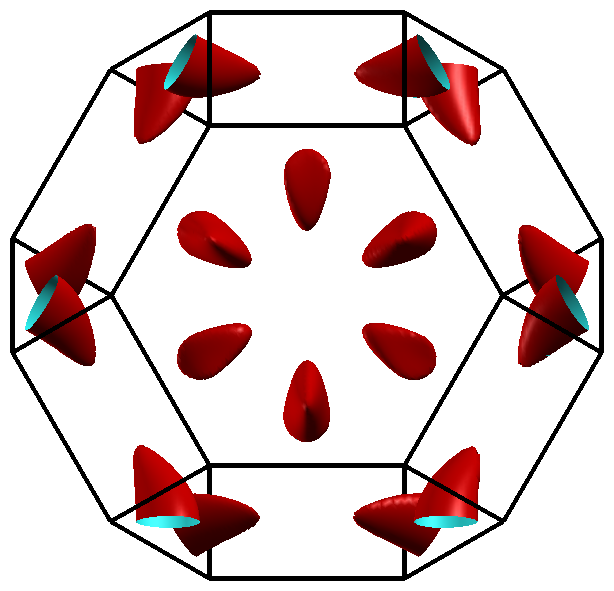}
			\end{subfigure}
			\caption{\textbf{Electronic structure of Co$_{\bf3}$Sn$_{\bf2}$S$_{\bf2}$.} \textbf{(a)} The spin-polarized (red for spin-up and blue for spin-down channel) and SOC-included (dotted green curves) band structure plotted along the high symmetry direction  W$-$T$-$U$-$L$-$$\Gamma$. The spin-down channel has a gap, whereas the spin-up channel is gapless and has two band-crossings along U$-$L$-$$\Gamma$ which are parts of a nodal ring. \textbf{(b)} The orbital-projected density of states. The large area under red curve implies that the Co-\textit{d} orbital dominates near Fermi level with negligible contributions from other orbitals. The $ + $ve and $ - $ve values correspond to spin-up and spin-down channels respectively.  \textbf{(c)} The Brillouin zone of the system with a truncated octahedral shape. The green line shows high-symmetry direction for the band structure plot. \textbf{(d)}-\textbf{(e)} Fermi surface for the valence and conduction bands respectively. Both the bands cross Fermi level.}
			\label{fig:electronic_structure}
		\end{figure*}
	
	In this section, we investigate the electronic band structure of Co$_3$Sn$_2$S$_2$ in detail, both in the absence and in the presence of spin-orbit coupling (SOC). In the absence of SOC, the calculated electronic band structure, along the path W$-$T$-$U$-$L$-$$\Gamma$, is shown in Fig. \ref{fig:electronic_structure}(a). The corresponding high-symmetry points in the truncated octahedron-shaped Brillouin zone is shown in Fig. \ref{fig:electronic_structure}(c) with the green line indicating the high-symmetry path for the band structure. By calculating the spin-polarized band structure along the aforementioned path, we find (Fig. \ref{fig:electronic_structure}(a)) that the spin-down channel (blue) is insulating with a gap of 0.44 eV, whereas the spin-up channel (red) is gapless confirming its half-metallic nature. One of the main interest in the band structure lies in the two crossings along the path U$-$L$-$$\Gamma$. Energetically they are located slightly above and below the Fermi level. As the interaction between two spin-channels is avoided in this particular case, the mirror-symmetry of the Hamiltonian remains intact and leads to a nodal ring situated in the mirror plane. It turns out that these two crossings are parts of that particular nodal ring itself. Interestingly, due to the other two symmetries, namely the inversion symmetry and $C_{3z} $, six such nodal rings in total are present in the entire Brillouin zone.
	\begin{table}[hb]
		\begin{tabular}{|c|c|c|c|c|}
			\hline
			\bm{\quad $ k_1 $ \quad} & \bm{\quad $ k_2 $ \quad} & \bm{\quad $ k_3 $ \quad} & \bm{$ E-E_f  (meV)$} & \bm{$ Chirality $} \\ \hline
			$ \ \ \ 0.000 \ $ & $ \ \ \ 0.425 \ $ & $ \ \ \ 0.063 \ $ & $ 62.9 $ & $ +1 $ \\ \hline
			$ \ \ \ 0.000 \ $ & $      -0.425 \ $ & $      -0.063 \ $ & $ 62.5 $ & $ -1 $ \\ \hline
			$ \ \ \ 0.425 \ $ & $ \ \ \ 0.000 \ $ & $ \ \ \ 0.063 \ $ & $ 62.3 $ & $ +1 $ \\ \hline
			$      -0.425 \ $ & $ \ \ \ 0.000 \ $ & $      -0.063 \ $ & $ 62.6 $ & $ -1 $ \\ \hline
			$      -0.425 \ $ & $      -0.425 \ $ & $      -0.365 \ $ & $ 62.7 $ & $ +1 $ \\ \hline
			$ \ \ \ 0.425 \ $ & $ \ \ \ 0.425 \ $ & $ \ \ \ 0.365 \ $ & $ 62.5 $ & $ -1 $ \\ \hline
		\end{tabular}
		\caption{The locations (fractional coordinates) of the Weyl nodes in the Brillouin zone. Their energies above Fermi level and the chiralities are also shown in last two columns.}
		\label{tab:weyl_points}
	\end{table}

		\begin{figure}[!hbt]
			\includegraphics[width=8cm]{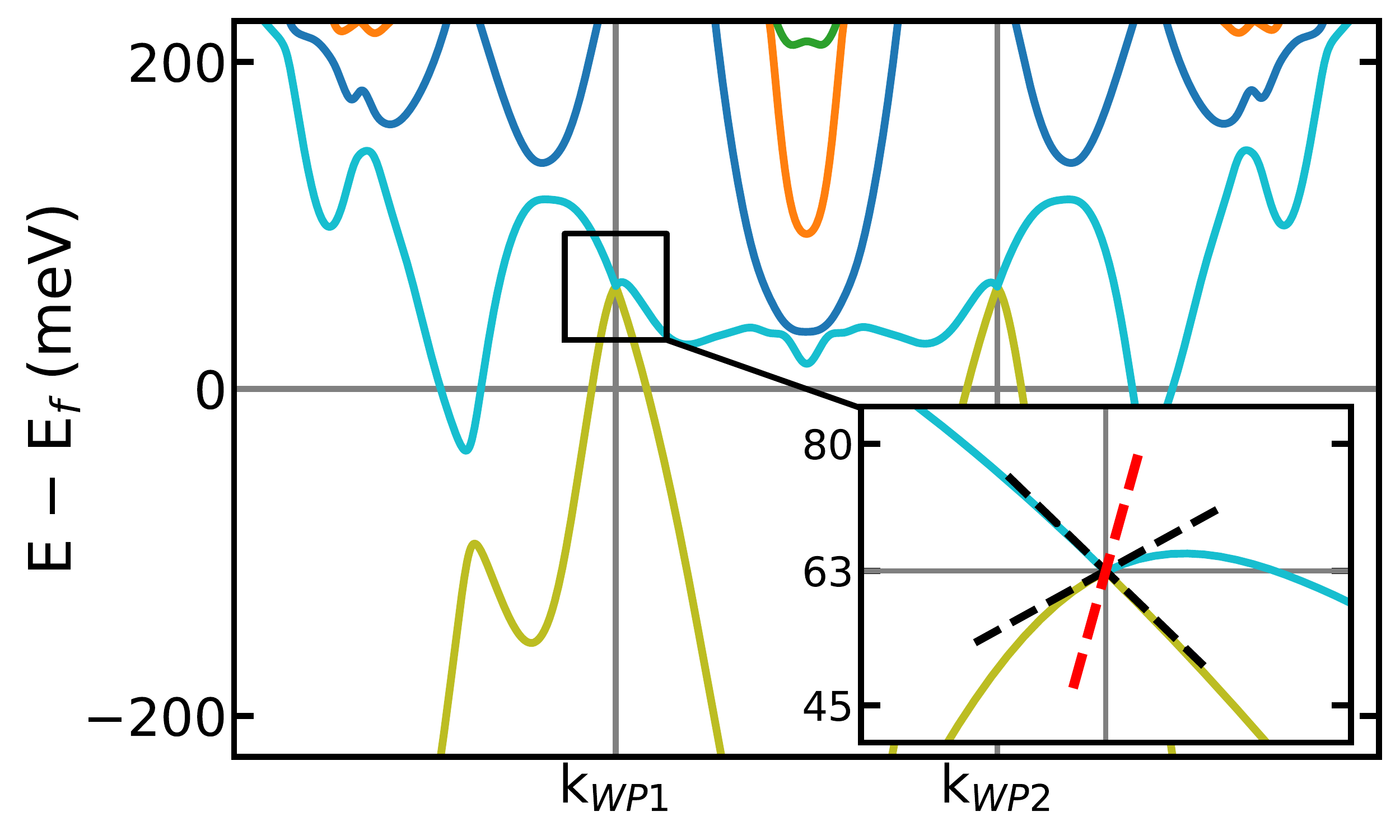}
			\caption{\textbf{Tilted Weyl cones.} Band structure showing Weyl nodes of opposite chiralities is plotted along their connecting line. Their locations being $ (0, 0.425, 0.063) $ and $ (0, -0.425, -0.063) $ respectively in fractional coordinates. One of them has been zoomed-in to show the tilting of the cone with respect to vertical line via the red line.} 
			\label{fig:weyl-cone}
		\end{figure}

	As the compound consists of elements with relatively large atomic numbers, it is desirable to incorporate spin-orbit interactions (which, typically, vary as the fourth power of the effective nuclear charge $Z$). The mirror-symmetry of the Hamiltonian gets broken with the introduction of SOC and hence the nodal ring, that was present earlier, collapses into a pair of Weyl points with opposite chiralities. The electronic band structure, including SOC, is shown with the dotted green curves in Fig. \ref{fig:electronic_structure}(a). The aforementioned band crossings around the Fermi level are gapped out, which essentially indicates that the Weyl nodes are not located on the high-symmetry line. Upon scanning the first Brillouin zone, we found six such Weyl nodes in total of alternate chiralities which happens due to the extant inversion and $C_{3z}$ crystal symmetries. All the nodes happen to be at around 63 meV above the Fermi level owing to the presence of TRS. Their locations in the Brillouin zone as well as corresponding energies and chiralities are listed in Table \ref{tab:weyl_points}. They are very close to Fermi level and hence we expect a significant contribution of the Weyl nodes to the transport.

	We have plotted the first two Weyl nodes of Tab. \ref{tab:weyl_points} in Fig. \ref{fig:weyl-cone}. Upon zooming in, it is found that the cones are tilted with respect to the vertical. To investigate this quantitatively, we consider an effective two-band Hamiltonian for a tilted WSM which is written as,
		\begin{equation}\label{eq:tilted_hamilt}
			H = v \ \bm{k.\sigma} + \bm{t.k} + \epsilon_0
		\end{equation}
	where $ \bm k $ is the momentum relative to the Weyl node. $ v $, $ \bm t $, and $ \epsilon_0 $ being the Fermi velocity, tilt parameter, and Weyl node energy respectively. The cones shown in Fig. \ref{fig:weyl-cone} are tilted primarily along the $ k_y $ direction. From our DFT calculation, we obtain the value of $ t_y $ and $ v $ as 0.19 and 0.65 eV$ \cdot\AA $. The ratio $ |\frac{t_y}{v}| $ turns out to be around 0.29 ($ < $ 1) which indicates that the WSM is of type-I. In a similar fashion, the tilt in other two pairs can also be shown, and they are along different directions. It is important to note that our calculations were done in lattice model \cite{gorbar_anomalous_2017,sharma_nernst_2016} with 54 bands rather than this two-band linearized model.

	The Co atoms consist of partially filled \textit{d}-orbital and, therefore, the electronic correlations are expected to play an important role in this system. Recently, Xu \textit{et al.} \cite{xu_electronic_2020}, have estimated the Coulomb interaction to be around 4 eV in this material by combining optical spectroscopy measurements with many-body theoretical calculations. We have tried to incorporate the Coulomb interaction in our calculations following the simplified (rotationally invariant) approach, given by Dudarev \textit{et al.} \cite{dudarev_electronenergyloss_1998}. However, it made the Weyl nodes disappear by gapping out the band crossings. This implies that the single-electron picture is not sufficient to reproduce the whole physics. Since we are only interested in features caused by the Weyl nodes (coming from the crossing of valence and conduction bands), we ignore the effects of correlation in the foregoing. We have also plotted the Fermi surfaces for the valence and conduction bands responsible for the Weyl nodes in Fig. \ref{fig:electronic_structure}(d) and \ref{fig:electronic_structure}(e) respectively. It is no surprise that both the bands have contributions to the Fermi surface, which is also visible in the band structure. To get an insight into the orbital contributions, we have computed the orbital-projected, spin-polarized density of states around $ E_f $. It is clearly visible from Fig. \ref{fig:electronic_structure}(b) that the major contribution comes from the \textit{d}-orbital of Co. As we move away from $ E_f $, contributions from the \textit{p}-orbitals of Sn and S atoms begin to arise. We can also see that the DOS profile of spin-up (positive values) and spin-down (negative values) channels are not exactly same. And that inequality is owed to the ferromagnetism, coming predominantly from the $3d$-orbital of Co. 
	
\section{Anomalous Thermal Hall Effect}\label{sec:athe}

	In this section, we will investigate the anomalous thermal	Hall effect using the structural and electronic properties of Co$_3$Sn$_2$S$_2$. The ATHE refers to the appearance of a transverse heat current as a first-order response to an applied longitudinal thermal gradient. In linear response regime, the equation for the heat current in the presence of applied thermal gradient ($\bm \nabla T$) and electric field ($\bm E$) is given by,
		\begin{equation}\label{eq:heat_curr}
			\bm J_Q = T \alpha . \bm E - \kappa . \bm \nabla{T}
		\end{equation}
	where $ \alpha $ and $ \kappa $ in Eq. \ref{eq:heat_curr} are the Peltier and thermal conductivity tensors corresponding to Nernst-Ettinghausen and Leduc-Righi effects respectively. The transverse component of these tensors can be expressed as \cite{xiao_berryphase_2006,bergman_theory_2010,mccormick_semiclassical_2017} $ \alpha_{xy} = \frac{k_B e}{h} c_1 , \ \kappa_{xy} = - \frac{k_B^2 T}{h} c_2 $ where,
		\begin{equation}
			c_i = \int \frac{d \bm k}{(2\pi)^3} \sum_{n} \Omega_{\textbf{k}n} \int_{\epsilon_{\bm k n} - \mu}^\infty d\epsilon_{\bm k n} (\beta \epsilon_{\bm k n})^i \dfrac{\partial f_{eq}(\epsilon_{\bm k n})}{\partial \epsilon_{\bm k n}}
		\end{equation} 
	with $ \Omega_{\textbf{k}n} $, $ \epsilon_{\bm k n} $, $ \mu $ and $ f_{eq} $ being the Berry curvature, energy eigenvalue of the \textit{n}-th band, chemical potential and the equilibrium Fermi distribution function respectively. $ \beta $ is written in place of $ 1/(k_B T) $, where $ k_B $ and $ T $ are the Boltzmann constant and temperature respectively. Upon simplifying the above equation, we can write the expression for anomalous thermal Hall coefficient as
		\begin{multline}\label{eq:athc}
			\kappa_{xy} = \frac{k_B^2 T}{h} \int \frac{d \bm k}{(2\pi)^3} \sum_{n} \Omega_{\textbf{k}n} [ \frac{\pi^2}{3} + \beta^2 (\epsilon_{\bm k n} - \mu)^2 f(\epsilon_{\bm k n} - \mu) \\
			- [ln(1+e^{-\beta (\epsilon_{\bm k n} - \mu)})]^2  - 2Li_2[1-f(\epsilon_{\bm k n} - \mu)] ]
		\end{multline}
	where $ Li_2(z) $ is the polylogarithm function of order 2.
		\begin{figure*}[ht]
			\centering
			\hspace{0.15cm}
			\begin{subfigure}{0.45\textwidth}
				\caption{}
				\label{sfig:berry_path}
				\includegraphics[width=\textwidth]{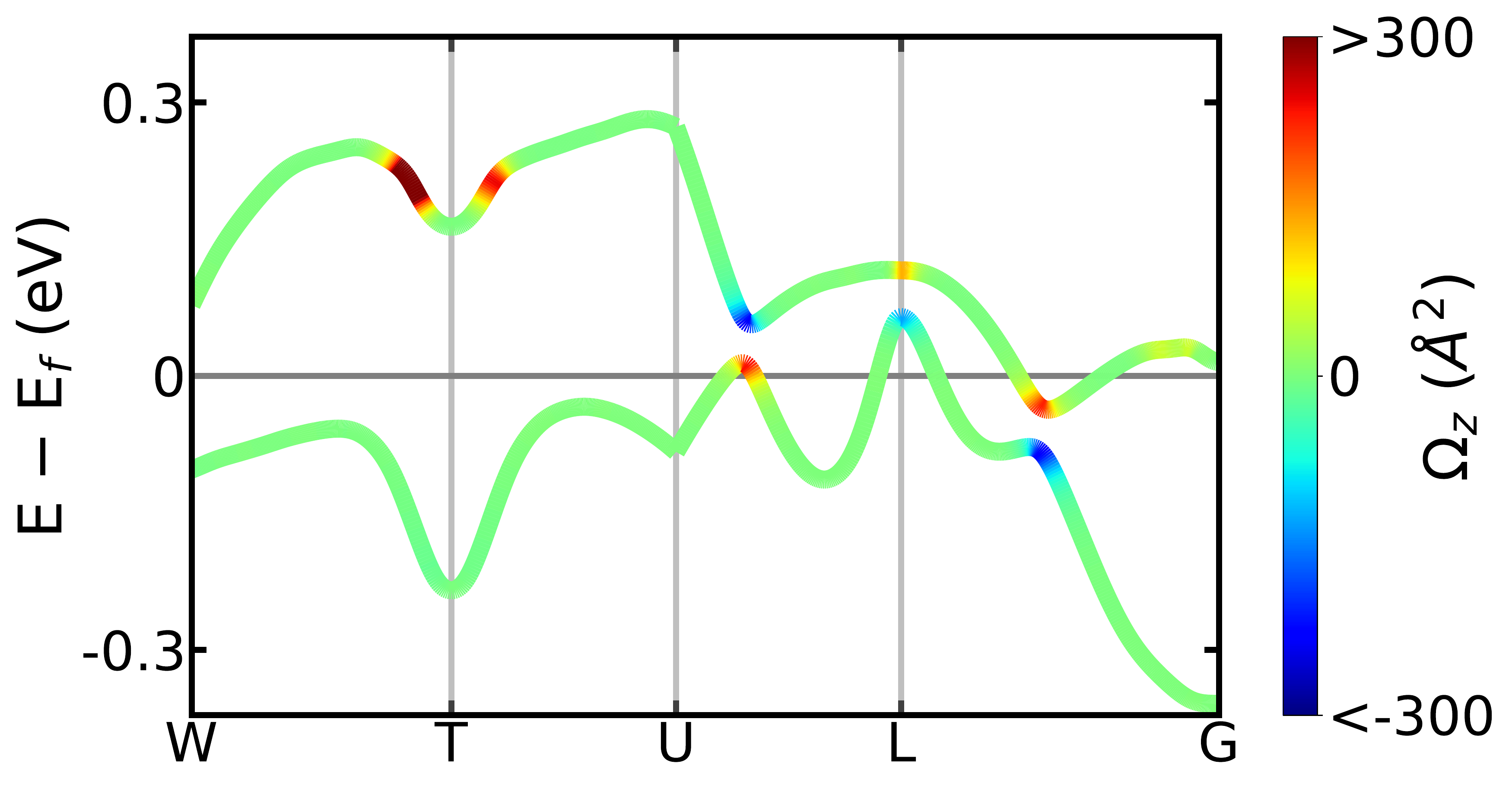}
			\end{subfigure} \hspace{0.8cm}
			\begin{subfigure}{0.338\textwidth}
				\caption{}
				\label{sfig:berry_plane}
				\includegraphics[width=\textwidth]{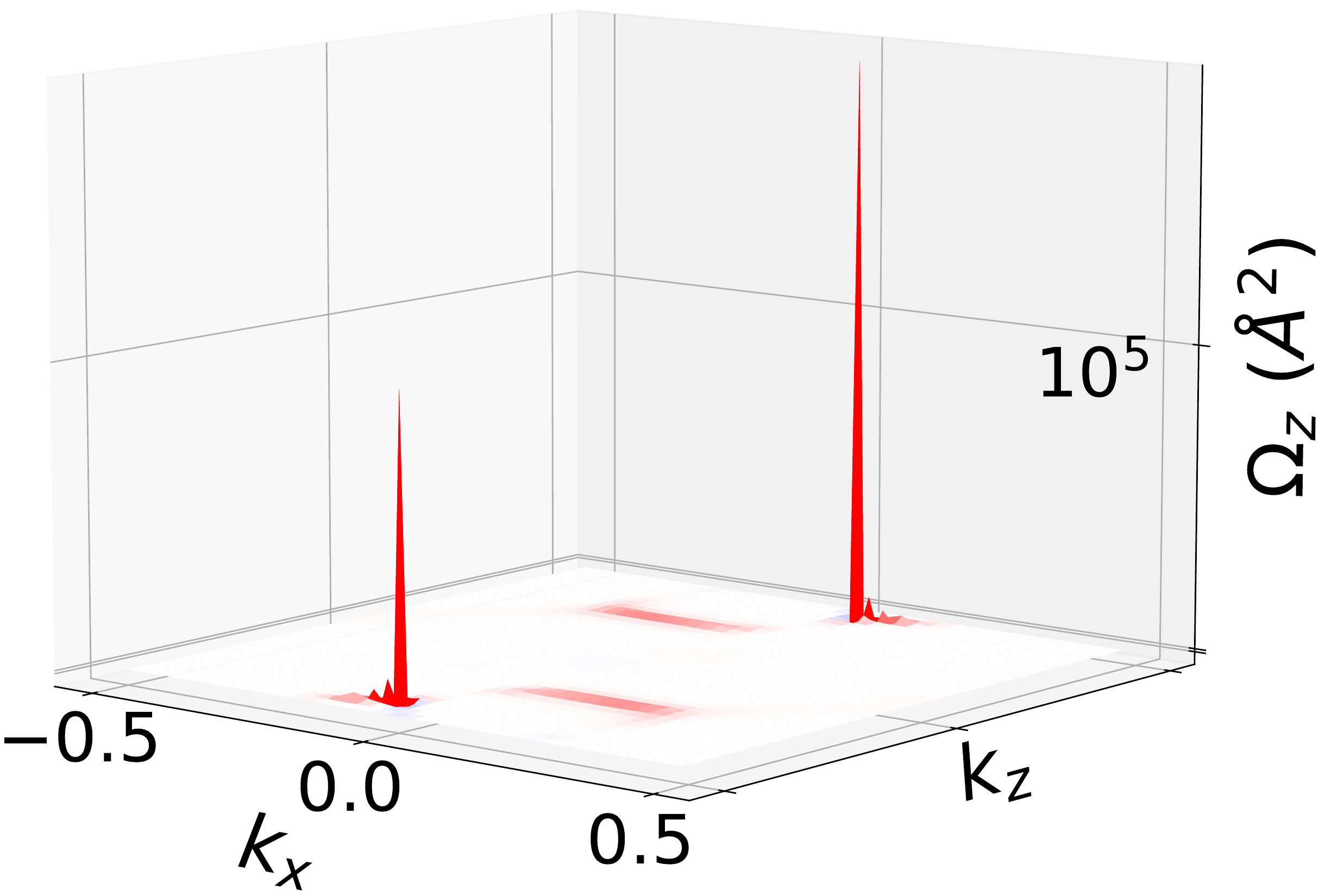}
			\end{subfigure}  \\ \vspace{0.3cm}
			\begin{subfigure}{0.40\textwidth}
				\caption{}
				\label{sfig:athc}
				\includegraphics[width=\textwidth]{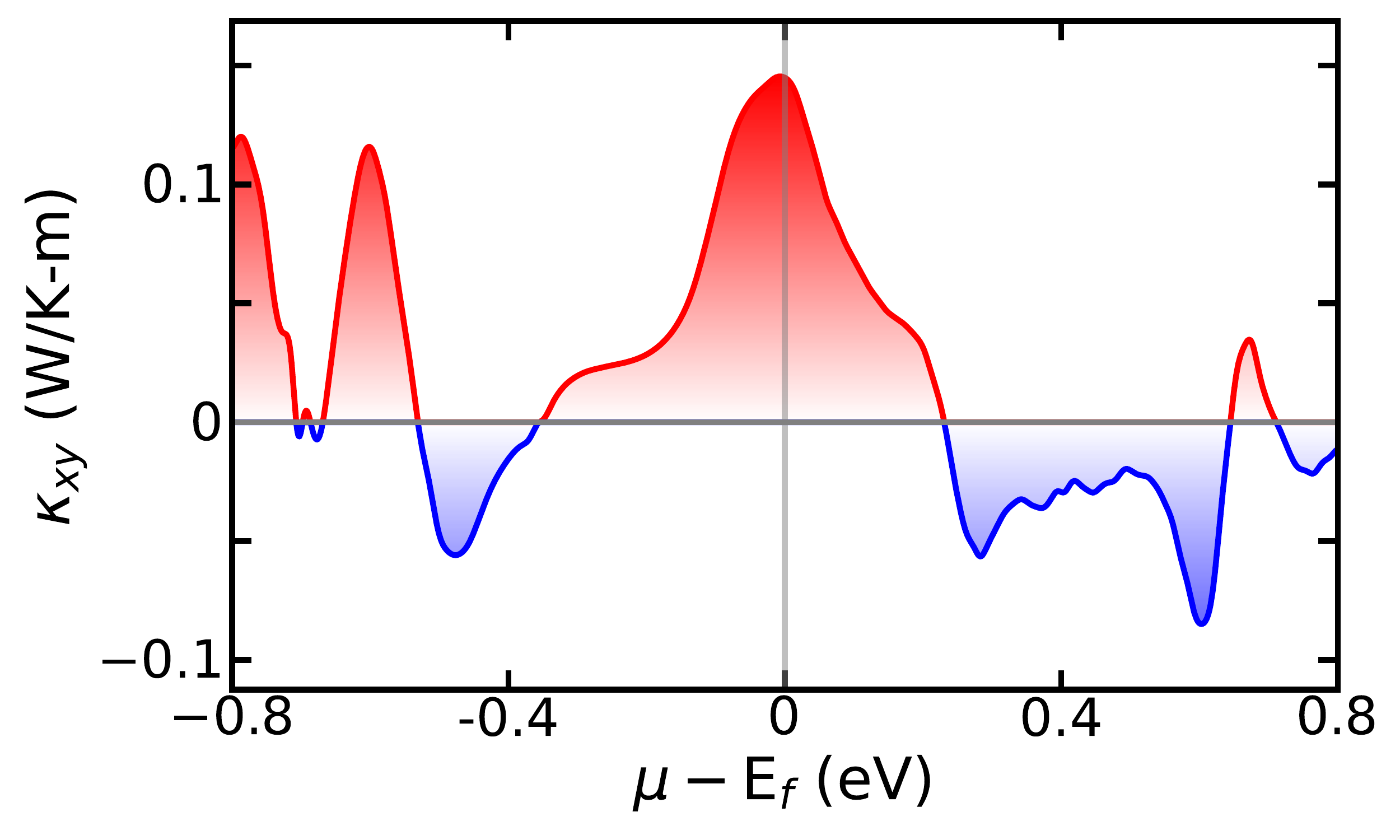}
			\end{subfigure} \hspace{0.8cm}
			\begin{subfigure}{0.40\textwidth}
				\caption{}
				\label{sfig:athc_Temp}
				\includegraphics[width=\textwidth]{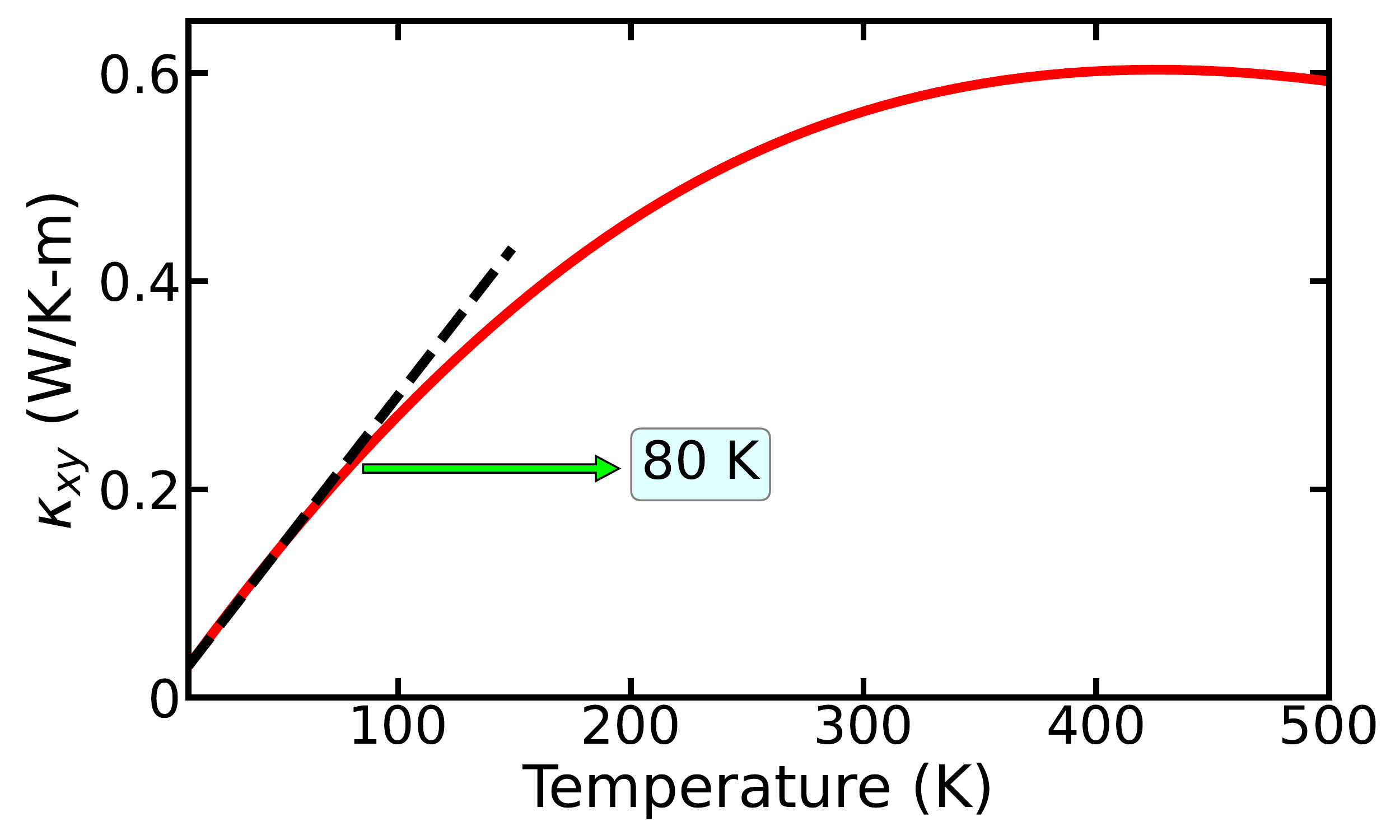}
			\end{subfigure}
			\caption{ \textbf{Berry curvature and anomalous thermal Hall conductivity.} \textbf{(a)} The strengths of the Berry curvature (\textit{z}-component) corresponding to the bands forming Weyl nodes are shown by color variation of the bands along high-symmetry path W$-$T$-$U$-$L$-$$\Gamma$. \textbf{(b)} The total Berry curvature is shown on the mirror-symmetric plane ($ k_y = 0 $). The sharp peaks correspond to the Weyl nodes present on this plane. \textbf{(c)} Anomalous thermal Hall conductivity (ATHC) plotted with respect to chemical potential $ \mu $ at $ T=50 $ K. The heat current can be controlled as well as reversed by changing the chemical potential of the system. However, the maximum value is achieved at $ \mu = $ E$_f$. \textbf{(d)} The relationship between ATHC and temperature. Note that $ \kappa_{xy} $ varies linearly with $ T $ at low temperatures up to 80 K after which it gradually deviates from linearity with increasing T. The chemical potential was set equal to the Fermi level.} 
			\label{fig:berry_athc}
		\end{figure*}

	It is important to note that we are considering only the transverse response to the applied thermal gradient $\nabla T$ in presence of nontrivial Berry curvature without any external magnetic field.
	In order to calculate ATHE, we must discuss first the Berry curvature profile of the system as we are considering only the Berry curvature-induced contribution. Since the TRS is broken in Co$_3$Sn$_2$S$_2$ due to its inherent ferromagnetism, it possesses finite values of Berry curvature throughout the Brillouin zone and sharp peaks at the locations of the Weyl nodes (Tab. \ref{tab:weyl_points}). The \textit{z}-component of the Berry curvature ($ \Omega_z $) along the high symmetry path (W$-$T$-$U$-$L$-$$\Gamma$) as well as on the mirror-plane (k$_x$-k$_z$) are depicted in Figures \ref{fig:berry_athc}(a) and \ref{fig:berry_athc}(b) respectively. 
	The values in the former are significantly lower than those in the latter because the Weyl nodes have dislocated from the high-symmetry line after activating the spin-orbit coupling.

	The plane plot, however, could capture the exact locations of the nodes as shown in Table \ref{tab:weyl_points} because the Berry curvature attains highest value at the Weyl points. The values at the momentum locations (0.425, 0, 0.0625) and (-0.425, 0, -0.0625) turn out to be in the range of 10$ ^5 $. After obtaining $\kappa_{xy}$ using Eq.~\ref{eq:athc}, we plot it as a function of chemical potential $ \mu $ in Fig. \ref{fig:berry_athc}(c). The magnitude of the ATHC is significantly large compared to other materials even in the absence of any external magnetic field \cite{hwang_topological_2020,li_topological_2021,ideue_giant_2017,zhang_anomalous_2021} and it reaches the maximum value of 0.146 W/K-m at its original Fermi level. In addition, we find that the direction of thermal Hall current can be reversed by tuning the chemical potential by, for example, doping the material. It is worth mentioning at this point that we have also performed a spin-polarized calculation of ATHC after switching off SOC where the material is not in Weyl semimetal phase. However, both spin channels produced null results at the original Fermi level.
		\begin{figure*}[!hbt]
			\centering
			\begin{subfigure}{0.33\textwidth}
				\caption{}
				\label{sfig:bands_strain-0.0}
				\includegraphics[width=\textwidth]{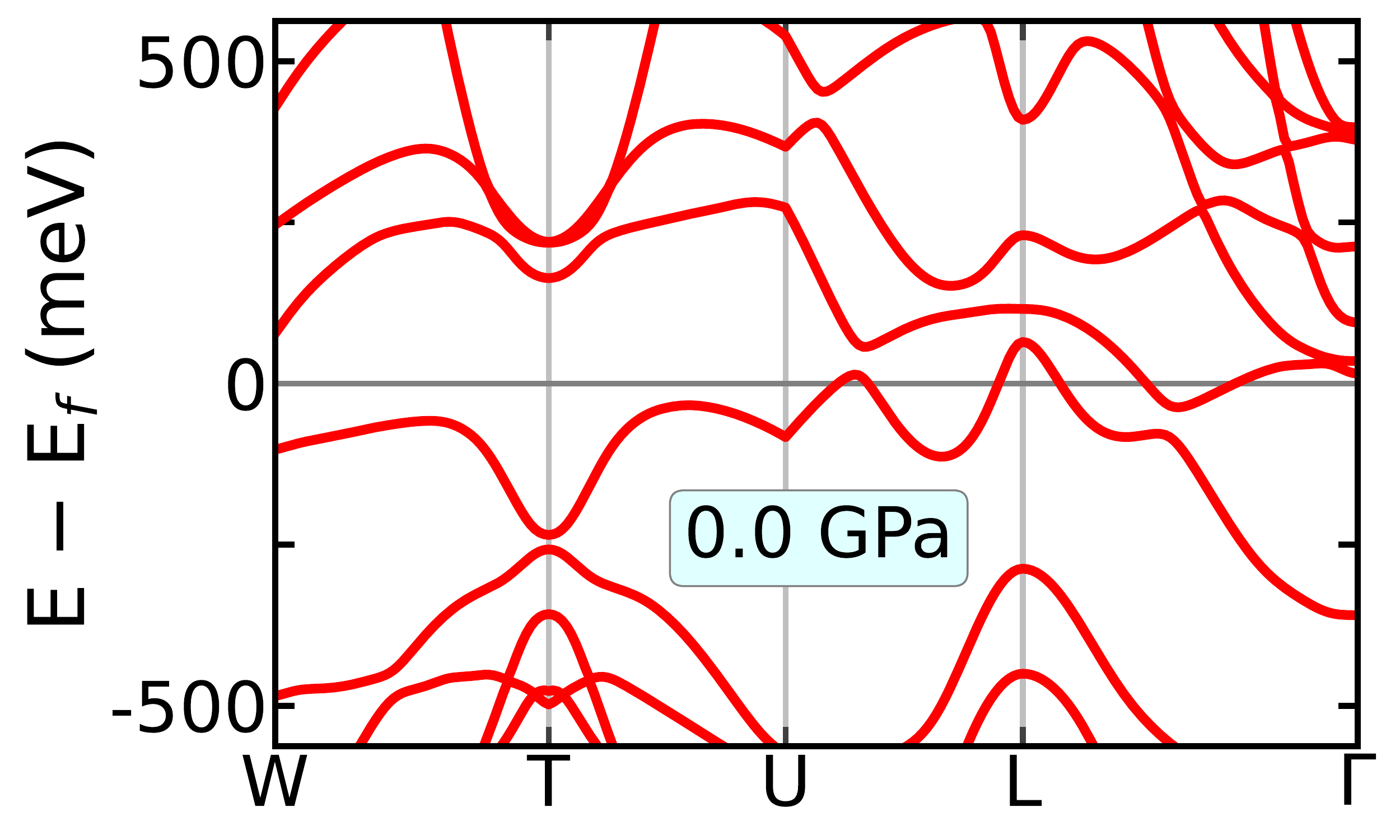}
			\end{subfigure} \hfill
			\begin{subfigure}{0.33\textwidth}
				\caption{}
				\label{sfig:bands_strain-2.5}
				\includegraphics[width=\textwidth]{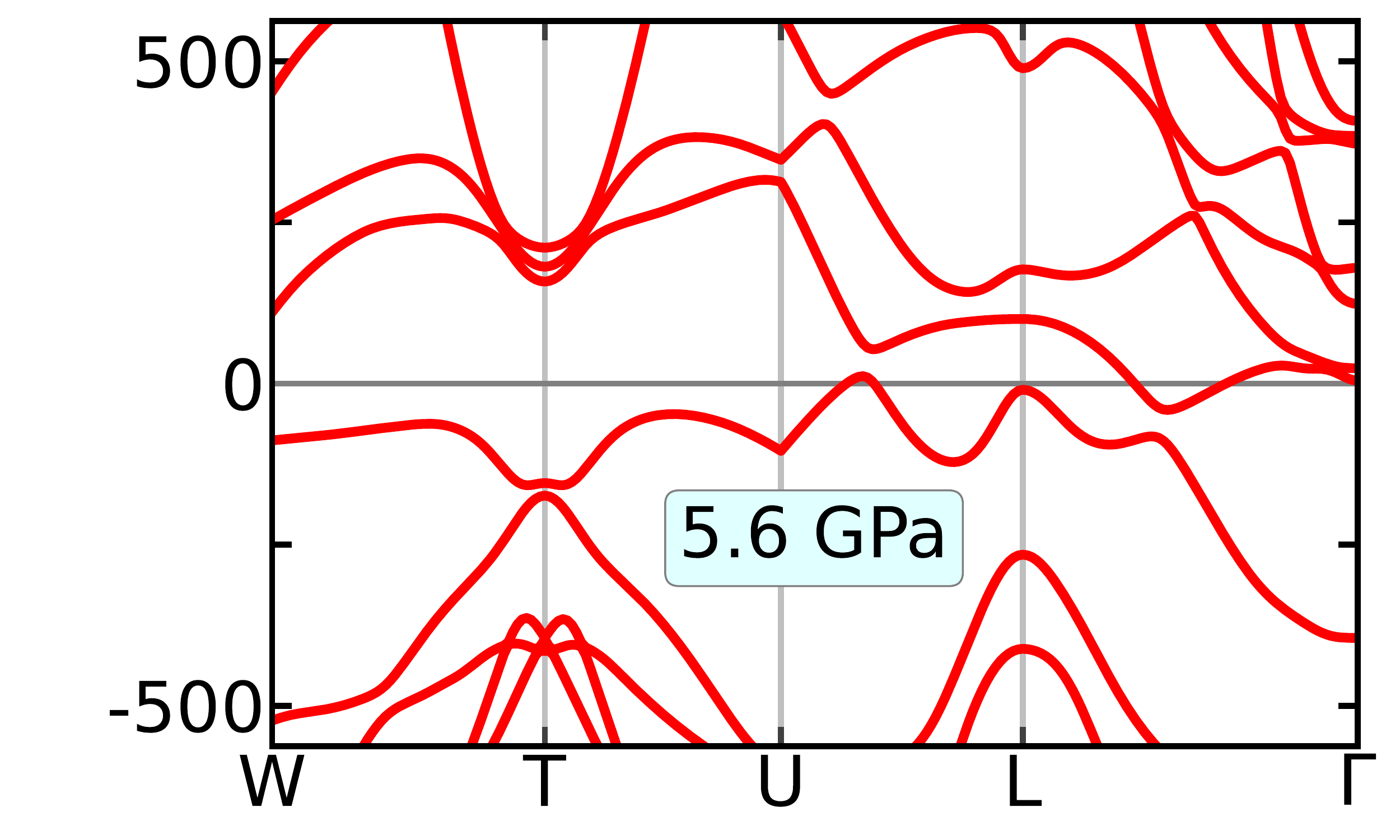}
			\end{subfigure} \hfill
			\begin{subfigure}{0.33\textwidth}
				\caption{}
				\label{sfig:bands_strain-5.0}
				\includegraphics[width=\textwidth]{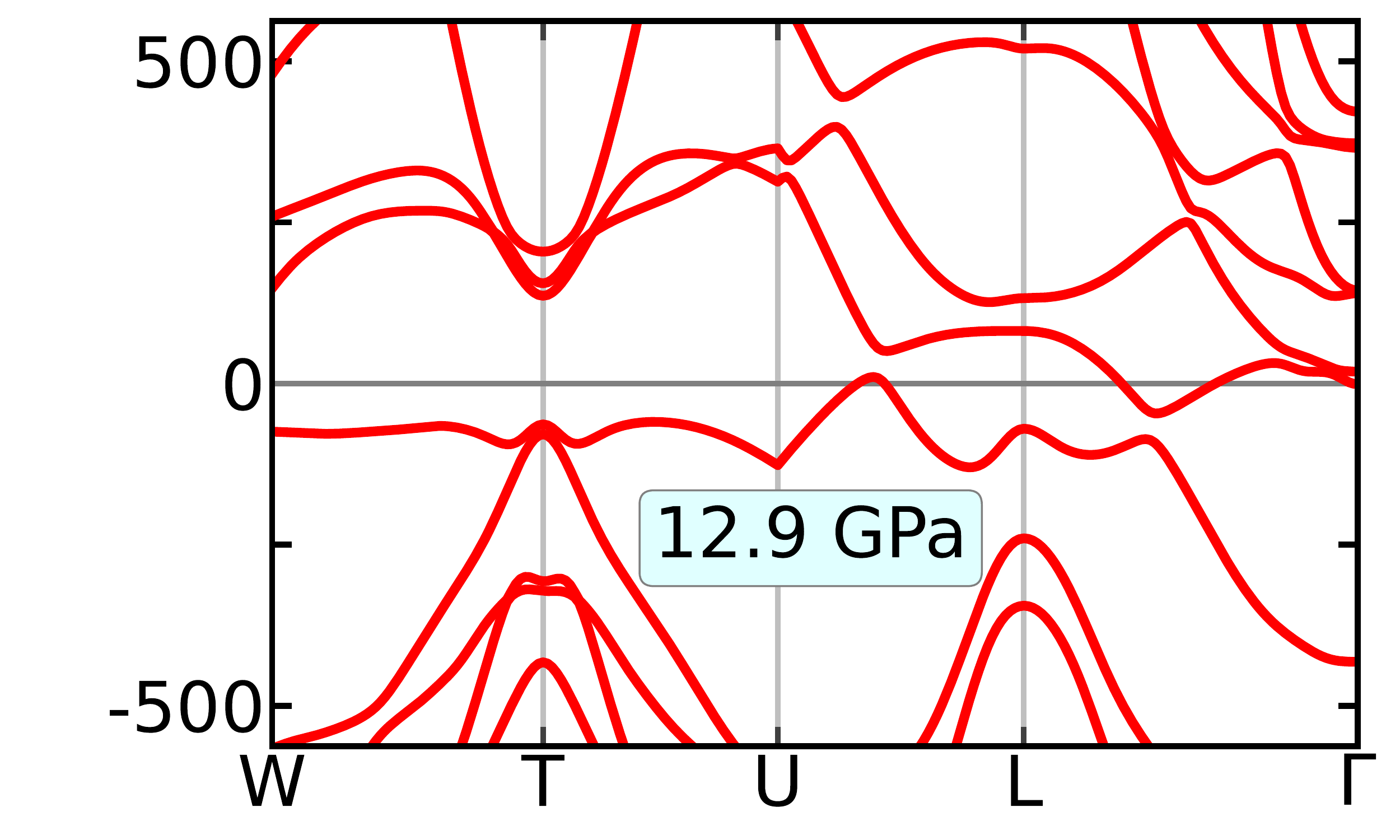}
			\end{subfigure} \\ \vspace{0.2cm}
			\begin{subfigure}{0.33\textwidth}
				\caption{}
				\label{sfig:nodes_strain-0.0}
				\includegraphics[width=\textwidth]{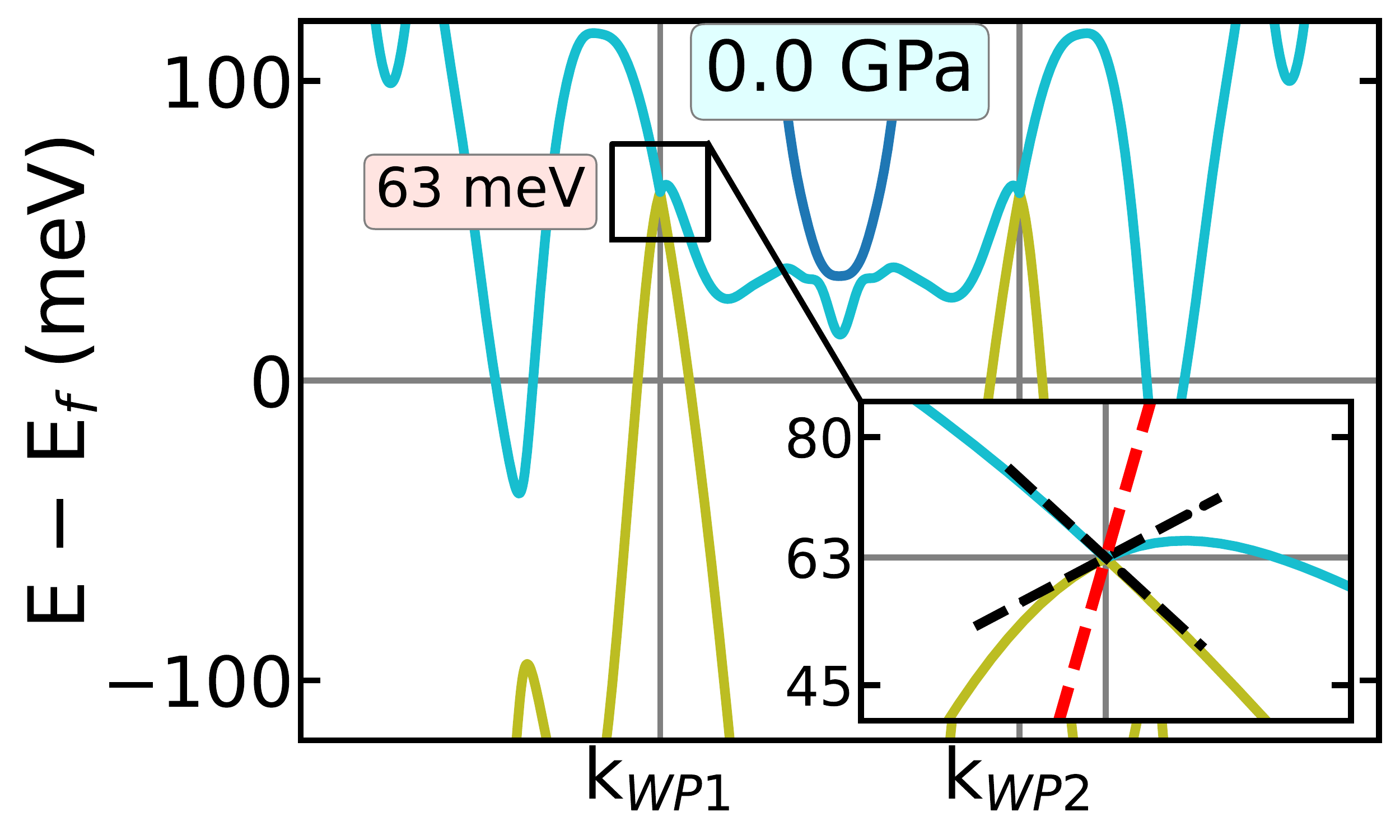}
			\end{subfigure} \hfill
			\begin{subfigure}{0.33\textwidth}
				\caption{}
				\label{sfig:nodes_strain-2.5}
				\includegraphics[width=\textwidth]{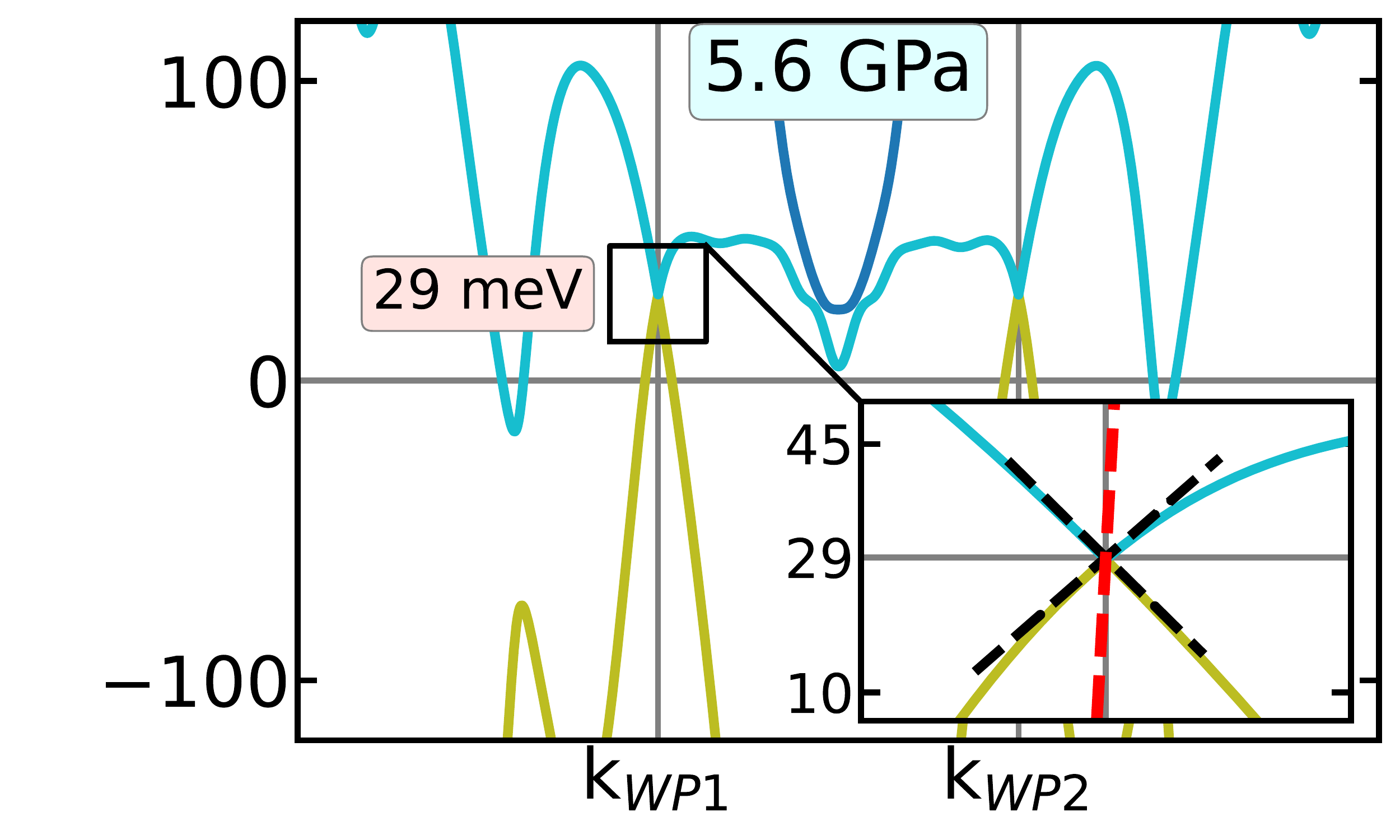}
			\end{subfigure} \hfill
			\begin{subfigure}{0.33\textwidth}
				\caption{}
				\label{sfig:nodes_strain-5.0}
				\includegraphics[width=\textwidth]{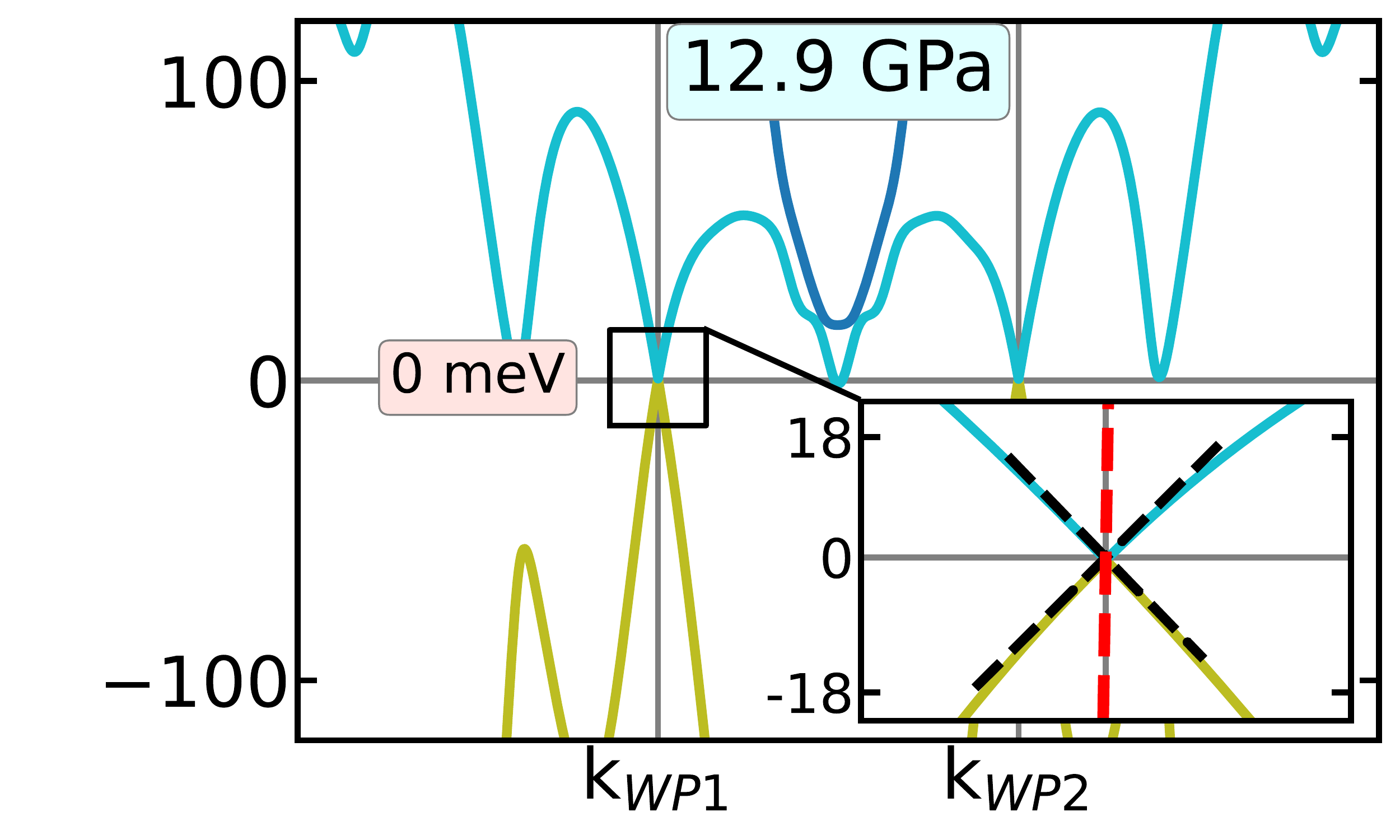}
			\end{subfigure}
			\caption{ \textbf{Strain-induced band dispersion.} \textbf{(a)}-\textbf{(c)} SOC-included band structures for compressive stresses of 0, 5.6, and 12.9 GPa along \textit{z}-axis. The valence band starts to flatten as the stress is increased. The local band gap at \textbf{L} point shows an expanding nature throughout the process. \textbf{(d)}-\textbf{(f)} The band dispersion around two of the Weyl nodes along their connecting line in the Brillouin zone. The nodes move towards the Fermi level with increasing stress and finally at 12.9 GPa they fall directly on the Fermi level. The tilting of the cones decreases with stress and almost vanish at 12.9 GPa. The red line in the inset indicates tilting with respect to the vertical. }
			\label{fig:strain_bands}
		\end{figure*}
	
	To understand the main source of this large effect we analyze all the terms in Eq. \ref{eq:athc}. As the system possesses inversion symmetry, the full integration of the Berry curvature over the Brillouin zone is zero. Hence, the first constant term of the kernel written inside the square bracket has no contribution. At low temperatures, the integration of second and third terms inside the square bracket turns out to be exactly equal and opposite. We would like to point out that these terms give a very large contribution at the high temperature limit. Hence, the only contribution to ATHC comes from the polylogarithm function multiplying with Berry curvature at $ T \rightarrow 0 $. It is important to note that the $x$ and $y$ components of the Berry curvature lead to a very small Hall conductivity (less than 5\% of that due to the \textit{z}-component) and hence only the \textit{xy}-component of ATHC has been shown.

	At low temperature, Eq. \ref{eq:athc} is reduced to
		\begin{equation}\label{eq:WF_law}
			\kappa_{xy} = \frac{\pi^2}{3}\frac{k_B^2 T}{h} \int \frac{d \bm k}{(2\pi)^3} \sum_{n} \Omega_n \theta (\mu - \epsilon_{\bm k n}) = L_0 T \sigma_{xy}
		\end{equation}
	where $ L_0 $ is known as the Lorenz number and has a value of $ 2.44 \times 10^{-8} \ W\Omega K^{-2} $. The Eq. \ref{eq:WF_law} satisfies the Wiedemann-Franz law which states that the ratio of the thermal conductivity and the electrical conductivity of the material is directly proportional to temperature. To validate this from our calculation, we have computed the anomalous Hall conductivity ($ \sigma_{xy} $) at low temperature and computed the ratio of $ \sigma_{xy} $ and $ \kappa_{xy} $. The proportionality constant comes out to be very close to $ L_0 $ over the range of chemical potential ($ \mu $) at 50 K. However, as T is increased, Eq. \ref{eq:WF_law} no longer remains valid.

	We plot $ \kappa_{xy} $ over a large range of temperature $ T $ in Fig. \ref{fig:berry_athc}(d). When $ T $ is low, $ \kappa_{xy} $ increases linearly with T up to 80 K according to the Wiedemann-Franz law. However, as the temperature is further increased, it slowly loses its linear character and after a certain temperature ($\sim$ 400K), it starts to decrease. This indicates that the Wiedemann-Franz law for anomalous transports is only valid at low temperatures. In the next section we see what happens to the transport properties when one applies external pressure on this system.

%

\section{Effects of strain}\label{sec:strain}

	In this section we shall investigate the effect of strain on the anomalous transport properties in Co$ _3 $Sn$ _2 $S$ _2 $. Here we restrict ourselves to only uniaxial compressive stress which is achieved by introducing strain in the lattice vector. For this purpose, we reduce the unit cell length along \textit{z}-direction gradually by up to 5\% which is equivalent to a stress of around 13 GPa along the \textit{z}-axis. At every step of increasing stress, the structure has been geometry optimized by keeping the cell volume fixed while relaxing the atomic positions. And it is important to note that no structural transition has taken place during the process. 
	\begin{figure}[hb]
		\centering
		\includegraphics[width=8cm]{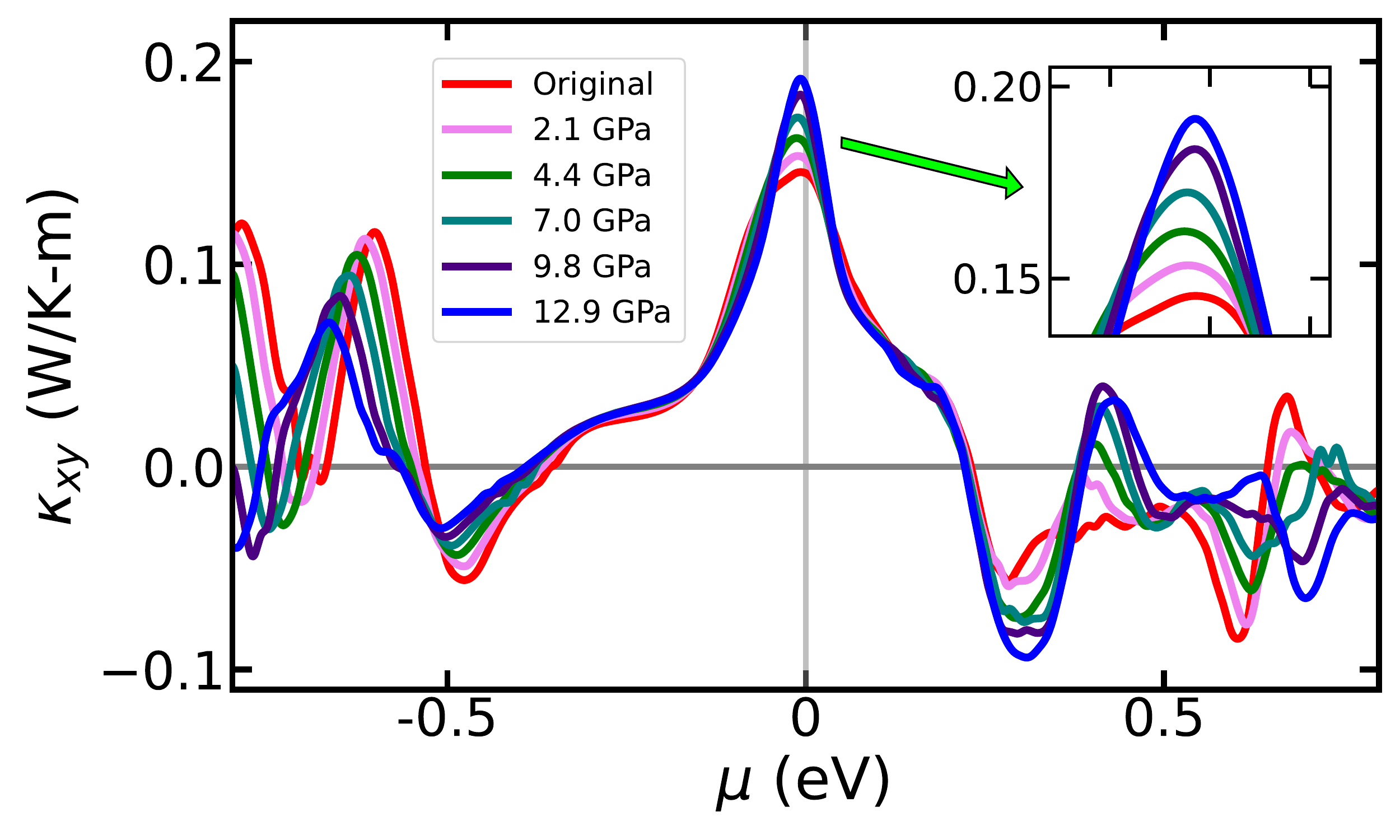}
		\caption{ The anomalous thermal Hall conductivity plotted against chemical potential for different values of compressive stress along \textit{z}-axis. At $ \mu = $ E$_f$, ATHC increases by about 33\% for a stress of 12.9 GPa.}
		\label{fig:athc_strain}
	\end{figure}

	To investigate what happens to the electronic properties, we plot the SOC-included band structure in Fig. \ref{fig:strain_bands} (upper row) for three different stresses. First thing we observe is that the Fermi level gets shifted away from the valence band (gapped nodal ring region) as the pressure is increased. In addition, the valence band flattens out in some regions of the k-path. As a consequence, the local band gap at $ L $ point also gets expanded. The band dispersion around the first pair of the Weyl nodes are also shown in the lower part of Fig. \ref{fig:strain_bands}. It is quite remarkable to observe that the Weyl nodes are being dragged towards the Fermi level as the stress is increased and at a stress of 12.9 GPa the nodes fall directly on the Fermi level which initially were at 63 meV above $ E_f $ in normal condition. Another interesting thing to note here is, the tilting of the Weyl nodes decreases with increasing stress. This can be observed in the insets in figures \ref{fig:strain_bands}(d)-(f) where the tilt-axis (red line) becomes almost vertical at 12.9 GPa. Quantitatively, the tilt ratio $ |\frac{t}{v}| $ becomes $ \sim 0.02 $ in Fig. \ref{fig:strain_bands}(f).

	As topological effects come from the bands nearest to the Fermi level, we expect that this movement of the Weyl points towards $ E_f $ should show up as an enhancement in the topological transport phenomena. In order to see that quantitatively, we plot anomalous thermal Hall conductivity for different values of external stress in Fig. \ref{fig:athc_strain}. Despite having almost same pattern, the values of ATHC do vary at some regions along $ \mu $ axis. At their original Fermi levels ATHC increases by about 33\% with just 5\% compressive strain, which is huge. But, in the left part of the figure ATHC is seen to decrease with increasing stress. However, since experimentally one measures the outcome from the original Fermi level, the conductivity will increase. We had also performed calculations with strain along lateral directions, but did not observe any enhancement. Hence, we conclude from this section that the effect of uniaxial compressive strain in this magnetic Weyl semimetal is quite significant which can enhance the anomalous thermal Hall conductivity drastically.

\section{Summary}\label{sec:summary}

	To summarize, we have investigated the Berry curvature-induced anomalous thermal Hall effect in  magnetic Weyl semimetal Co$ _3 $Sn$ _2 $S$ _2 $ using a combination of the first principles DFT calculation and quasi-classical Boltzmann transport theory. From DFT calculation, we first obtain the electronic band structure of the material and observe that the system transits from a nodal-line semimetal to a Weyl semimetal by turning on the spin-orbit coupling. We show that the first Brillouin zone contains three pairs of Weyl nodes in total, created by the $ C_{3z} $ rotational symmetry and they are located at the same energy due to the presence of inversion symmetry. We have shown from DFT calculations that the Weyl cones are tilted and also calculated the tilt parameter which indicates that the system is a type-I WSM. From a tight-binding Hamiltonian, derived from Wannier functions, we have calculated the Berry curvature which shows large enhancement especially at the positions of the Weyl nodes.
	
	Next, using the semiclassical Boltzmann transport theory, we have computed the Berry curvature-induced anomalous thermal Hall conductivity of Co$ _3 $Sn$ _2 $S$ _2 $ which came out to be very large. By varying the chemical potential the thermal Hall current can be tuned and even the direction can be reversed for some particular values. We have also checked the Wiedemann-Franz law and found that this law is well satisfied at low temperatures. In addition, to further enhance the thermal conductivity in Co$ _3 $Sn$ _2 $S$ _2 $, we apply uniaxial compressive stress. This essentially moves the Weyl nodes towards the Fermi level which in turn enhances ATHC as the anomalous transport properties are directly related to the Weyl nodes. To be quantitative, the conductivity increased by about 33\% for a compressive strain of just 5\% along the \textit{z}-axis of the unit cell. It is worthwhile to mention that the tilt of the Weyl cones vanishes along the connecting line on application of stress.
	
	We would like to point out that since the \textit{d}-orbitals of Co atom contribute the most to the bands near Fermi level which create the Weyl nodes and these orbitals are partially filled, one can expect a non-trivial role of electronic correlation in this material. However, the incorporation of Coulomb interaction in DFT (i.e., DFT+U) gaps out the Weyl node crossing which is unphysical and, therefore, one has to go beyond the single-electron Hartree-Fock picture to fully understand the effects of correlation in this material.

\section*{Acknowledgments}
	
	A.R.K. acknowledges National Supercomputing Mission (NSM) for providing computing resources of `PARAM Shakti' at IIT Kharagpur, which is implemented by C-DAC and supported by the Ministry of Electronics and Information Technology (MeitY) and Department of Science and Technology (DST), Government of India.

	\bibliography{citations}
\end{document}